\documentclass[apj]{emulateapj}
\usepackage{graphicx}

\def      \mnras        {MNRAS}
\def       \apj          {ApJ}
\def       \apjs         {ApJS}
\def       \apjl         {ApJ}
\def       \aj           {AJ}
\def       \aap          {A\&A}
\def       \aaps         {A\&AS}
\def       \araa         {ARA\&A}
\def       \pasj         {PASJ}

\shorttitle{Dust destruction in non-radiative shocks}
\shortauthors{Zhu et al.}

\begin{document}

\title{
Dust destruction in non-radiative shocks
}

\author{H.~Zhu$^{1,2}$, P.~Slane$^{2}$, J. Raymond$^{2}$, W.W.~Tian$^{1,3,4}$}
%\tiny{
\affil{$^1$Key Laboratory of Optical Astronomy,
    National Astronomical Observatories,
    Chinese Academy of Sciences,\\
    Beijing 100012, China;
    zhuhui@bao.ac.cn, tww@bao.ac.cn}
\affil{$^2$Harvard-Smithsonian Center for Astrophysics,
    60 Garden Street, Cambridge, MA 02138, USA;\\
    pslane@cfa.harvard.edu, jraymond@cfa.harvard.edu}
\affil{$^3$School of Astronomy and Space Science, University of Chinese Academy of Sciences,
    Beijing 100049, China}
\affil{$^4$Department of Physics and Astronomy,
    University of Calgary,
    Calgary, Alberta T2N 1N4, Canada}
%    }

\begin{abstract}
Supernova remnant (SNR) shock waves are the main place where
interstellar dust grains are destroyed. However, the dust destruction
efficiency in non-radiative shocks is still not well known. One way
to estimate the fraction of dust destroyed is to compare the
difference between postshock gas abundances and preshock medium
total abundances when the preshock elemental depletion factors are
known. We compare the postshock gas abundances of 16 SNRs in Large
Magellanic Cloud (LMC) with the LMC interstellar medium abundances
that we derived based on 69 slow-rotating early B-type stars. We
find that, on average, $\sim$61\% of Si rich dust grains are destroyed
in the shock while the fraction of dust destroyed is only $\sim$40\%
for Fe rich dust grains. This result supports the idea that
the high depletion of Fe in the diffuse neutral medium is not caused by the resilience
of Fe rich grains but because of faster growth rate. This work also
presents a potential way to constrain the chemical composition of
interstellar dust.
\end{abstract}


\section{Introduction}
Interstellar dust is a fundamental component of galaxies (Draine
2011). It modulates the spectra of galaxies by absorbing short
wavelength radiation and re-radiating the energy in the infrared
band. The thermal emission of dust could be an important cooling
mechanism of collapsing dense molecular clouds, allowing star
formation to occur. Dust also provides the formation site of H$_2$
molecules and shields molecules from ultraviolet radiation, which
makes it central to interstellar gas chemistry. Dust even plays a
critical role in controlling the temperature of the diffuse
interstellar medium (ISM) by locking heavy elements inside it and
ejecting photoelectrons into the gas. \\

It has been confirmed that dust can form: 1) in the ejecta of
supernovae; 2) in the outflows of evolved stars (e.g. the red giant
stars, carbon stars, planetary nebulae), and 3) through regrowth
in the ISM. The first formation route is considered as an explanation
for large amounts of dust seen in high redshift galaxies (e.g.,
Mailiano et al. 2004; Dwek et al. 2007; Valiante et al. 2009; Gall
et al. 2011), while the second route contributes dust only to
``present-day'' galaxies which are old enough for low and intermediate
mass stars to leave the main sequence (e.g., Draine 2003). Usually,
dust from these two routes is called stardust. Technically, the regrowth
route should work for any galaxies with sufficiently high densities to allow
metal elements to accrete onto dust grain surfaces. The unsolved question
is how much dust is from the regrowth process (e.g., Draine 2009;
Dwek \& Cherchneff 2011; Temim et al. 2015; Slavin et al. 2015).
Studying dust destruction in the shocks of supernova remnants (SNRs)
could help approach this issue because any imbalance between dust
formation of the first two routes and dust destruction from SNR shock
waves will necessitate dust regrowth in the ISM. For example,
a higher dust destruction rate will bring into question the ejecta
of supernovae as the main dust formation site of high redshift galaxies
and support the idea that supernovae may only provide the dust seeds to grow in the ISM.\\

During the past few decades, numerous studies have been done on
dust destruction in shocks (e.g., Shull 1977; Draine \& Salpeter
1979; McKee et al. 1987; Dwek et al. 1996; Jones et al. 1996;
Williams et al. 2006; Borkowski et al. 2006; Raymond et al. 2013;
Bochico et al. 2014; Slavin et al. 2015; Laki{\'c}evi{\'c} et al. 2015; Dopita et al. 2016, 2018).
Previous studies suggested that dust destruction is usually caused
by thermal/non-thermal sputtering, shattering and vaporization.
Sputtering dominates over other processes in fast non-radiative
shocks. The main processes are non-thermal sputtering and shattering
in slow radiative shocks. However, the rate of dust destruction is
still poorly known. Table 1 lists the measured fraction of dust
destroyed in several SNRs in the Milky Way and Large Magellanic
Cloud (LMC). As can be seen, the fraction of dust destroyed in
non-radiative shocks varies from $\sim$20\% to $\sim$50\% with an
averaged value of 34\%. For radiative shocks, it could rise to more
than 50\% (Dopita et al. 2016, 2018). Compared with the recent
numerical hydrodynamical model of Slavin et al. (2015), the measured
fraction of dust destroyed in non-radiative shocks is lower than
the prediction (see their Figure 7). \\

In this paper, we measure an averaged fraction of dust destroyed
by non-radiative shocks based on the postshock gas abundances of
16 LMC SNRs. We then apply the result to constrain recent dust
models. In section 2, we describe the general concept and equation
to calculate the fraction of dust destroyed. In section 3, we show
how to obtain the values for the three key parameters. The results
and discussion are presented in section 4. \\

\begin{table*}[!htbp]
  \centering
  \caption{The fraction of dust destroyed for 8 SNRs with non-radiative shock.}
    \begin{tabular}{@{}p{3cm}p{2cm}p{2cm}p{3cm}p{3.5cm}@{}}%{lcccr}
    %\begin{tabular}{lcccr}
    \hline
    \hline
    Name        & Location  & Age (yr)   & Percent destroyed & Reference \\
    \hline
    Cygnus Loop & Milky Way & 10000      & 35\%             & Sankrit et al. 2010 \\
    Puppis A    & Milky Way & 3700       & 25\%             & Arendt et al. 2010 \\
    N49B        & LMC       & 10900      & 27\%             & Williams et al. 2006 \\
    0453-68.5   & LMC       & 8700       & 33\%             & Williams et al. 2006 \\
    N23         & LMC       & 4600       & 39\%             & Williams et al. 2006 \\
    N132D       & LMC       & 2500       & 38\% - 50\%      & Williams et al. 2006 \\
    DEM L71     & LMC       & 4400       & 35\%             & Borkowski et al. 2006 \\
    0548-70.4   & LMC       & 7100       & 40\%             & Borkowski et al. 2006 \\
    \hline
    \end{tabular}%
    %\end{tabular}%
  \label{tab:tab1}%
\end{table*}%

\section{The general concept and equation}

The X-ray radiation of young and middle-aged SNRs is usually dominated
by thermal emission from forward-shocked ISM and reverse-shocked
ejecta which are separated by a contact discontinuity. Under
strong shock conditions, the thickness of the shocked ISM is
about $R/12$ with ${\rm R}$ as the radius of the SNR.  This is
sufficiently large that, for a typical SNR, it is possible to measure
the elemental abundances of the shocked ISM without any pollution from
shocked ejecta with a high spatial-resolution telescope, e.g. {\it
Chandra}. \\

Suppose one element, X, in the preshock ISM has an abundance of
${X}_{\rm LMC}$. The measured postshock gas abundance of element X,
from the shock front to contact discontinuity,
is ${X}_{\rm po},$
and the depletion factor of this element, ${D_{\rm X}}$,
is defined as the logarithm of its reduction factor below the
expected abundance relative to that of hydrogen if all of the atoms
were in the gas phase (Jenkins 2009):
\begin{equation}
{D_{\rm X}} = {\rm Log[N(X)/N(H)]} - {\rm Log(X/H)_{\rm LMC}},
\end{equation}
in which ${\rm N(X)}$ is the column density of X in gas
phase and ${\rm N(H)}$ is the column density of hydrogen. Finally,
we have the expression:
\begin{equation}
X_{\rm po} = X_{\rm LMC} \times 10^{D_{\rm X}} + R_{\rm X} \times \left(1 - 10^{D_{\rm X}}\right) \times X_{\rm LMC}.
\end{equation}
At the right side of Equation 2, the first part is the preshock gas
abundance for element X and the second part presents the abundance
for the same element returned to gas phase.

The fraction of dust
destroyed, $R_{\rm X}$, is given by
\begin{equation}
R_{\rm X} = \frac{X_{\rm po} - 10^{D_{\rm X}} \times X_{\rm LMC}}{X_{\rm LMC} \times {\rm (1-{10}}^{D_{\rm X}})}.
\end{equation}
As is shown above, we can obtain ${R_{\rm X}}$ if we have
the values for ${X}_{\rm po}$, ${X}_{\rm LMC}$ and ${D_{\rm X}}$.
In the next section, we will introduce how to measure these key
parameters.\\

\section{Three key parameters}
\subsection{The postshock gas abundances, $X_{\rm po}$}

The postshock gas abundances, $X_{\rm po}$, are taken from Schenck
et al. (2016) where the authors carried out an abundance study of
16 LMC SNRs. By making a three-color map and investigating the
literature for each SNR, they selected several small and/or thin
regions in the outermost boundary of each SNR for which X-ray
radiation is from the shocked ISM. The regions used for spectral
fits are shown in their Figure 1. Each region contains at least
$\sim$ 3000 counts to allow statistically significant spectral model
fits. A nonequilibrium ionization plane-parallel shock model, {\it
vpshock}, with two foreground absorption components from the Milky
Way and LMC, is used to fit the X-ray spectra. They then averaged
these measured abundances for each SNR and listed them in their
Table 2, with 90\% confidence level uncertainties. It's
worth noting that, in the definition of $X_{\rm po}$, the postshock
region is defined as the region from shock front to the contact
discontinuity. The $'$three-color map$'$ method used by Schenck et
al. (2016) should naturally guarantee that the regions in their
paper contain the whole postshock region. In cases where it might
not, the measured
$X_{\rm po}$ value will be less than the true value and $R_{\rm X}$
will be underestimated according to equation (3).\\

\subsection{The ISM abundances of LMC, $X_{\rm LMC}$}

The chemical composition of {\sc Hii} regions has long been used
to reference ISM abundances. However, previous studies demonstrated
that several heavy elements such as Mg, Si, and Fe show significant
depletions in {\sc Hii} regions (Esteban et al.  1998). Another
long-standing unsolved issue for {\sc Hii} regions is that the
derived abundances depend on both the fluctuations of the electron
temperature throughout the nebula and the lines used in the analysis,
i.e. the recombination or collisionally excited lines (Peimbert 2005).
These make the abundances of {\sc Hii} regions unsuitable for our
study. Red giant branch stars can also be used to study ISM
abundances. Unfortunately, they belong to the old stellar population and
therefore are not good tracers of present-day ISM abundances (Lapenna
et al. 2012). An alternative class of objects to provide ISM
abundances is that of the slowly-rotating early B-type stars. Unlike
{\sc Hii} regions, their composition is not affected by elemental
depletions. Considering the fact that they are young and have
clean photospheres without strong stellar winds or pollution from
convection (no evidence was found for effects of rotational mixing
up to projected rotational velocities of 130 km s$^{-1}$, e.g. Korn
et al. 2005), the B-type stars form an ideal reference for present-day
ISM abundances. \\

Since little or no abundance variation between cluster
members and field stars has been found (e.g. Korn et al. 2000;
Rolleston et al. 2002; Hunter et al. 2007, 2009; Turndle et al.
2007), we use both cluster and field B-type stars to reference the
LMC ISM abundances. To do this, we collected 69 B-type stars with
abundance measurements from different parts of the LMC (6 from Korn et
al. 2000, 4 from Korn et al. 2002, 3 from Rolleston et al. 2002, 3
from Korn et al. 2005, 30 from Hunter et al. 2007 and 23 from Turndle
et al. 2007). All of them have projected stellar rotational velocities
less than 150 km s$^{-1}$, indicating nearly all of them are
slow-rotating stars if the view is not pole-on. In Rolleston et al.
(2002), Hunter et al. (2007) and Trundle et al. (2007), the Si
abundances measured based on lines from different ionization states
are listed separately. Therefore, we calculate the uncertainty-weighted
mean Si abundance for each star by:

\begin{equation}
X_{\rm LMC} = \frac{\sum_{i=1}\left(W_iX_i\right)}{\sum_{i=1}W_i},
\end{equation}
with the weight
\begin{equation}
W_i = \frac{1}{\sigma_i^2}
\end{equation}
where ${X_{\rm i}}$ and ${\sigma_{\rm i}}$ are the Si abundance and
the associated uncertainty for each ionization state. The uncertainty
of the mean is
\begin{equation}
\sigma = \sqrt{\frac{\sum_{i=1}\left(W_i\sigma_i\right)^2}{\left(\sum_{i=1}{W_i}\right)^2}}
\end{equation}

The above three equations are also used to estimate the uncertainty-
weighted present-day global ISM chemical abundances of the LMC and we
list the result in Table 2.\\

\begin{table}[!htbp]
  \begin{center}
  \normalsize
  \caption{ISM abundances of LMC referenced by slow-rotating early B-type stars and the averaged postshock gas abundances of 16 LMC SNRs.}
    \begin{tabular}{lccr}
    \hline
    \hline
    Elements & \multicolumn{2}{c}{Abundance [X/H]$^a$}  \\
          & LMC             & SNRs            \\
    \hline
    C     & 7.62 $\pm$ 0.05 & -               \\
    N     & 7.39 $\pm$ 0.03 & -               \\
    O     & 8.35 $\pm$ 0.04 & 8.04 $\pm$ 0.04 \\
    Mg    & 7.07 $\pm$ 0.04 & 6.88 $\pm$ 0.06 \\
    Si    & 7.13 $\pm$ 0.06 & 6.99 $\pm$ 0.11 \\
    Fe    & 7.21 $\pm$ 0.08 & 6.84 $\pm$ 0.05 \\
    \hline
    \end{tabular}%
    \end{center}
    $^a$: Abundances are expressed as [X/H] = 12 + Log(X/H).\\
    Notes: The confidence level is 90\%.
  \label{tab:tab2}%
\end{table}%

\subsection{The depletion factor, ${D_{\rm X}}$}

As shown in equation (1), the depletion factor, $D_{\rm
X}$, can be used to describe how much of element X is peeled off
from the gas form and locked up into other forms (mainly in the form
of dust grains). Previous works showed that different elements
usually have different $D_{\rm X}$ values, and the overall strengths
of depletions of many elements differ along different lines of sight
(e.g., Jenkins et al. 1986; Crinklaw et al. 1994). It has been also
noted that the depletion factor is correlated with the average
volume density of hydrogen along each sight line (e.g. Savage \&
Bohlin 1979). However, those works were carried out for each element
independently without a systematic treatment of the relative
depletions of the elements. Jenkins (2009) solved this problem by
introducing a new parameter called the depletion strength factor, ${F_*}$,
which represents how far the depletion processes have progressed collectively for all elements
in any given case. Then the depletion factor for different elements
can be given by a simple linear relation in a two-parameters form
\begin{equation}
D_{\rm X} = D_{\rm X,0} + A_{\rm X}F_*
\end{equation}
or a three-parameter form
\begin{equation}
D_{\rm X} = B_{\rm X} + A_{\rm X}(F_* - z_{\rm X}),
\end{equation}
where $D_{\rm X,0}$ and $A_{\rm X}$ in Equation (7) are the zero
point and slope respectively. In equation (8), the zero point reference
for $F_*$ is $z_{\rm X}$, which is unique to element $X$, and $B_{\rm X}$
is the depletion at that point. For the Milky Way, Jenkins (2009) also found that
${F_*}$ is linearly correlated with logarithmic volume density n(H)
\begin{equation}
F_* = 0.772 + 0.461 {\rm log\left \langle n(H)\right \rangle}
\end{equation}
which means we could give an estimate of the local ${D_{\rm X}}$
for each element if we knew the local volume density. Unfortunately,
for the LMC, there is no ${F_*}$-log~n(H) relation because the distance
scale that the LMC neutral gas occupies is unknown, making it nearly
impossible to convert column density to volume density. Therefore,
we can't estimate the $D_{\rm X}$ value for each individual LMC SNR even
if the local volume density can be derived by fitting their X-ray
spectra. But it is still possible to obtain an averaged $D_{\rm X}$
value for 16 SNRs.\\

The most recent knowledge about elemental depletions in the LMC
is from Tchernyshyov et al. (2015). We rechecked the star sample
used by these authors. According to the positions of the stars on
the LMC HI column density map (see their Figure 1) and the measured
gas column densities for each stars (see their Figure 7), we found
no obvious space-distribution bias
(e.g., preference for star formation regions or the most diffuse ISM)
in their sample.
Furthermore, the ISM surrounding LMC SNRs shows even higher dust-to-gas
ratios than the averaged ratio of the LMC (Temin et al.  2015).
Therefore, it should be safe and maybe somewhat conservative to
take the averaged depletion factors in Tchernyshyov et al. (2015)
as the elemental depletions for 16 LMC SNRs used in our study. \\

The depletion factor of an element depends on the ISM total abundance
(gas plus dust). Comparing our LMC ISM abundances with the abundances
used by Tchernyshyov et al. (2015), there are differences of 0.22
and 0.11 for Si and Fe respectively. We use both values to correct
the Si and Fe depletion factors in Tchernyshyov et al. (2015) and
obtain the averaged depletion factors of -0.60 and -1.28 for Si and
Fe. This indicates that, on average, 75\% of Si and 95\% of Fe are locked
in dust grains.\\

\section{Result and discussion}
\subsection{The averaged fraction of dust destroyed}

With the averaged depletion factors, LMC ISM abundances and postshock
abundances in hand, now we can calculate the fraction of dust
destroyed for the 16 SNRs with equation 3. The results are reported
in Table 3. As can be seen, the fraction of Si-rich dust grains
destroyed for each SNR is usually higher than the value for Fe
rich dust grains. The averaged fraction of destroyed Si-rich dust
grains is ${61}^{+16}_{-11}$\% while it is only ${40}^{+5}_{-1}$\%
for Fe rich dust grains. The 21\% difference is illustrated clearly
in Figure 1 with the 61\% destruction dashed line well above
most of the data points for Fe. \\

\begin{figure}[!htpb]
\centerline{\includegraphics[width=0.45\textwidth, angle=0]{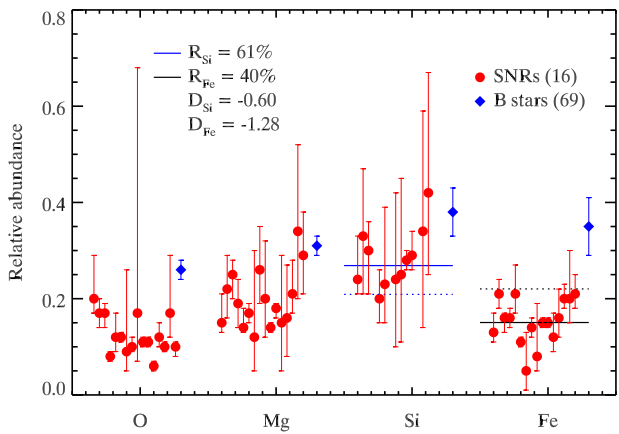}}
\caption{The comparison between the postshock gas abundances of 16 SNRs and
ISM abundances referenced by early B-type stars. The reference solar
abundances are taken from Anders \& Grevesse (1989). The red points show the measured postshock abundances of the 16 LMC SNRs and the blue diamonds are the LMC ISM abundances traced by 69 B-type stars. $D_{Si}$ and $D_{Fe}$ represent the depletion factors of Si and Fe. The $R_{Si}$ and $R_{Fe}$ stand for the averaged fraction of destroyed Si-rich and Fe-rich dust grains which are illustrated as solid blue and black lines respectively. In addition, the dotted blue and black lines mark the predicted postshock relative abundances assuming that 40\% Si-rich dust grains and 61\% Fe-rich dust grains are destroyed.}
\label{fig:fig1}
\end{figure}

\begin{figure}[!htpb]
\centerline{\includegraphics[width=0.45\textwidth, angle=0]{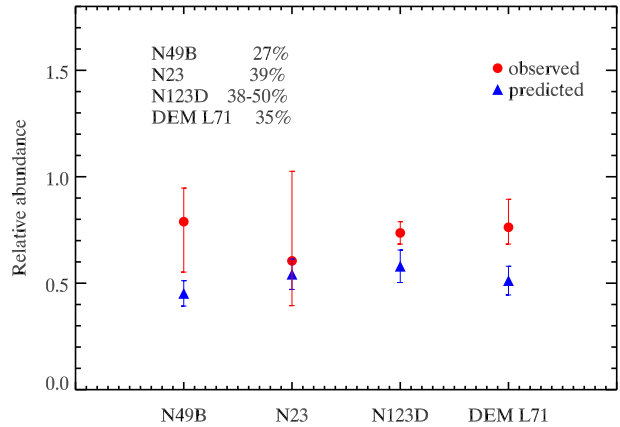}}
\caption{The comparison between predicted postshock Si abundances and measured Si abundances relative to LMC Si abundance referenced by slow rotating early B-type stars. The blue triangles are the predicted postshock abundances using previous published fraction of dust destroyed listed in Table 1. The remaining symbols and labels have the same meaning as Figure 1.}
\label{fig:fig2}
\end{figure}

Moreover, 4 SNRs in our sample have previous fractional dust
destruction measurements by Williams et al. (2006) and Borkowski
et al. (2006), as shown in Table 1. Adopting the previous value of
fractional destruction, we predict the postshock gas abundances and
compare them with the abundances from fitting the X-ray spectra in
Figure 2. The mismatch between the predictions and measurements for
Si are obvious which means that the fraction of dust destroyed that
is estimated in our work
is higher than previous ones. It is worth noting
that Williams et al. (2006) and Borkowski et al. (2006) found the
dust to gas mass ratio is lower by a factor of $\sim$4 than the
typical value calculated by Weingartner \& Draine (2001) for the
LMC.  Williams et al. (2006) explained this discrepancy by suggesting
that the dust grains are porous which can reduce the preshock dust
content and enhance the sputtering rate. Here, we demonstrate that,
with our result, the discrepancy can be explained naturally without
the assumption of porous dust grains. In the Weingartner \& Draine
(2001) dust model for the LMC, the Si abundance is assumed to be
1.67$\times 10^{-5} {\rm H}^{-1}$ in dust grains which lead to a
dust to gas mass ratio of 2.5$\times 10^{-3}$. However, the Si
abundance in dust grains is only 1.35$\times$0.75$\times 10^{-5}=1.01\times 10^{-5}
{\rm H}^{-1}$ referenced by B-type stars. If the abundances of other
elements are reduced by the same amount,
the dust to gas mass ratio will
decrease to $(1.01/1.67)\times$2.5$\times 10^{-3}=1.53\times 10^{-3}$. With the postshock
dust to gas mass ratio of about (1/4)$\times$2.5$\times 10^{-3}$ from Williams et al. (2006),
the fraction of destroyed dust will be 1-(0.25$\times$2.5$\times
10^{-3}$)/(0.61$\times$2.5$\times 10^{-3}$) = 0.6. This value is
consistent with our measurement of ${61}^{+16}_{-11}$\%.\\

\begin{table*}[!htbp]
  \begin{center}
  \caption{The fraction of dust destroyed of 16 LMC SNRs for Si-rich and Fe-rich dust grains.}
    \begin{tabular}{lcccc}
    \hline
    \hline
    Name       & Si                       & Percent destroyed & Fe                       & Percent destroyed \\
    SNR        &                          & \%                &                          & \% \\
    \hline
    N63A       & ${0.24}^{+0.09}_{-0.03}$ & 46                & ${0.13}^{+0.04}_{-0.02}$ & 36  \\
    N49        & ${0.33}^{+0.14}_{-0.12}$ & 81                & ${0.21}^{+0.03}_{-0.03}$ & 61  \\
    N49B       & ${0.30}^{+0.06}_{-0.09}$ & 69                & ${0.16}^{+0.01}_{-0.03}$ & 45  \\
    0453-68.5  &                          &                   & ${0.16}^{+0.02}_{-0.02}$ & 45  \\
    0540-69.3  & ${0.20}^{+0.06}_{-0.05}$ & 30                & ${0.21}^{+0.06}_{-0.04}$ & 61  \\
    N23        & ${0.23}^{+0.16}_{-0.08}$ & 42                & ${0.11}^{+0.01}_{-0.01}$ & 30  \\
    N157B      &                          &                   & ${0.05}^{+0.08}_{-0.04}$ & 11  \\
    N206       & ${0.24}^{+0.18}_{-0.14}$ & 46                & ${0.14}^{+0.02}_{-0.02}$ & 39  \\
    DEM L316B  & ${0.25}^{+0.20}_{-0.14}$ & 50                & ${0.08}^{+0.11}_{-0.03}$ & 20  \\
    N132D      & ${0.28}^{+0.02}_{-0.02}$ & 61                & ${0.15}^{+0.01}_{-0.01}$ & 42  \\
    DEM L71    & ${0.29}^{+0.05}_{-0.03}$ & 65                & ${0.15}^{+0.01}_{-0.01}$ & 42  \\
    DEM L238   &                          &                   & ${0.12}^{+0.05}_{-0.03}$ & 33  \\
    0519-69.0  & ${0.34}^{+0.25}_{-0.20}$ & 85                & ${0.16}^{+0.06}_{-0.04}$ & 45  \\
    0534-69.9  & ${0.42}^{+0.25}_{-0.17}$ & -                 & ${0.20}^{+0.03}_{-0.02}$ & 58  \\
    0548-70.4  &                          &                   & ${0.20}^{+0.10}_{-0.05}$ & 58  \\
    Honeycomb  &                          &                   & ${0.21}^{+0.04}_{-0.03}$ & 61  \\
    \hline
    Average    & ${0.28}^{+0.04}_{-0.03}$ & ${61}^{+16}_{-11}$ & ${0.15}^{+0.01}_{-0.01}$ & ${40}^{+5}_{-1}$ \\
    \hline
    \end{tabular}%
    \end{center}
    \textbf{Notes:} Column (2) and (4) are the abundances relative to solar (Anders \& Grevesse 1989).
  \label{tab:tab3}%
\end{table*}%

\subsection{Interstellar Fe and its depletion pattern}
The abundance of Fe is nearly comparable to Si and Mg (see Table 2).
This makes Fe an important component of interstellar dust, especially
for the warm neutral medium considering that Fe has a larger
depletion factor than Si and Mg. Interstellar olivine (Mg$_2x$Fe$_{2-2x}$SiO${_4}$) or
pyroxene (Mg$_x$Fe$_{1-x}$SiO${_3}$) provides possible candidates
for Fe (e.g., Mathis et al. 1977).  However, according to Poteet
et al. (2015), the silicate grains are Mg-rich rather than Fe-rich,
suggesting that most of the Fe is in other forms. Since the destroyed
fraction of dust inferred by Fe is smaller than the value from Si,
our work supports the Poteet et al.  (2015) result. The above
facts raise two questions. First: ``Where is the interstellar Fe?''
Potential reservoirs include iron oxides (Henning et al. 1995),
iron sulfides (Kohler et al. 2014), and metallic iron (Schalen 1965;
Draine \& Hensley 2013). Since insufficient amounts of iron oxides
and iron sulfides has been detected (Croat et al. 2005; Davis  2011)
and metallic iron has been found in different places of the Solar
System (Bradley 1994; Westphal et al. 2014; Altobelli et al. 2016),
metallic iron is the most likely candidate.\\

The second question about Fe is how to explain its depletion
pattern -- much higher depletion of Fe in the warm neutral medium
compared with Si. As shown in Figure 11 of Tchernyshyov (2015), the
solid-phase Fe abundance is roughly constant and always larger than
90\% of the total Fe abundance, while the solid-phase Si abundance
changes by a factor of about 2 from the least depleted case to the
most depleted case. Potential explanations are that the Fe-rich
dust grains are either (1)particularly resilient to survival in
interstellar shocks or (2)have a very fast regrowth rate even in
the diffuse ISM. \\

Jones et al. (2013) developed a new dust model with a
power-law distribution of small amorphous carbon grains and log-normal
distributions of large amorphous silicate and carbon grains, which
can naturally explain the infrared to far-ultraviolet extinction
and other observable dust emission and absorption features. In their
model, the answer to the first question is that 70\% of Fe exists
as the metallic inclusions of amorphous silicate grains. Zhukovska
et al. (2018) adopted this value in their analytic dust evolution
model which includes different residence times of dust grains and
dependence of dust growth on the density and temperature of the gas.
By further assuming the remaining 30\% Fe is in the form of free-flying
metallic nanoparticles, along with a steady state between dust destruction
and production by stellar sources and dust regrowth in ISM, they
found that the second question can be answered if the 70\% metallic
inclusion Fe is protected from rapid shock destruction. A direct
inference of the Zhukovska et al.  (2018) model is that the fraction
of Fe-rich dust destruction should be much smaller than 30\% because
70\% of Fe dust are protected by the silicate grains from shock
destruction. If we assume the fraction of destroyed free-flying
Fe nanopaticles has the same value as Si-rich dust grains, then the
actual fraction of Fe-rich dust destruction is $\sim$18\% which is
less than half of the measured value of ${40}^{+5}_{-1}$\%. Therefore,
our result supports that the high depletion of Fe in diffuse neutral
medium is not caused by the resilience of Fe rich grains, but its
fast growth even in diffuse ISM.\\

\subsection{Comparison with the abundances of O and Mg}
The results of Poteet et al. (2015) indicate that nearly all of Mg
is in the silicate dust grains in the line of sight of $\zeta$ Oph.
If it is also true for the LMC diffuse ISM, we expect that the amount of Si
and Mg returned to gas phase should be similar. Unfortunately, we
can not find the depletion factor of Mg in the literature to calculate
the amount of Mg returned to gas phase by equation ${X}_{po} -
X_{LMC} \times 10^{D_X}$. If we assume the amount of Mg released
from the dust grains is 70\% of the value of Si (Poteet et al.
2015), then we need $\sim$26\% of Mg in the gas phase for the
preshock ISM. This is a reasonable value if the depletion patterns
(the slope) of Mg are similar in both Milky Way and LMC. \\

The problem arises from the oxygen. Assuming the silicate dust is
in the form of olivine or pyroxene, then the amount of released O
should be 3 or 4 times larger than Si. If all the solid O is in silicate
dust grains, we need more than $\sim$62\% of O in solid phase which
is much higher than the predicted value (Jenkins 2009; Draine 2015).
The discrepancy may be explained in the light of:

(1) Solid O is not all in the form of silicate grains. Micrometer-sized
H${_2}$O ice grains could be an important reservoir of the missing
O without violating the observational constraints (Wang et al.
2015). However the destruction timescale of H${_2}$O ice
grains is short, (n$_e$t$\sim$10$^7$ cm$^{-3}$ s), which makes
H${_2}$O an inappropriate explanation of the O postshock
abundances. Another form to contain part of the missing O may be
Fe oxides.

(2) The measured postshock O abundances are not correct. Elemental
abundances are determined by line intensities. Strong lines
usually have more weight than weak lines during the determination.
Resonance line emission is caused by the transition between the
ground state and the first energy level of an ion. Resonance line
photons can be effectively scattered out of a special line of sight (LOS) to
another direction if the ion column density along that LOS is large.
Since all the regions used to determine postshock elemental
abundances are the limbs of the SNR, the LOS path length is much larger than
the path length in the perpendicular direction. Therefore, resonance line
scattering can lead to an underestimate of elemental abundances during
X-ray spectral fitting. X-ray spectroscopic observations of the Cygnus
Loop show that the relative abundance of O is two times smaller than
other heavy elements. Miyata et al. (2008) found evidence for resonance
line scattering for O VII K$\alpha$ line in the {\it Suzaku} X-ray spectrum of
the Cygnus Loop which can explain 20\% to 40\% of the difference.
So, we need to check whether resonance line scattering is
important or not in our sample. Taking N49B as an example, the
angular size is 2.6" which is equal to a diameter of 38 pc assuming
the distance of LMC is 50 kpc. The temperature is 0.52 keV while
the ionization timescale is 2.1$\times 10^{11} {\rm cm}^{-1} {\rm
s}$. Following Miyata et al. (2008), the line-center cross section
of resonance scattering can be expressed as
\begin{equation}
{\rm \sigma} = 1.86\times10^{-9}\frac{f}{E}{\rm \upsilon^{-1} cm^2},
\end{equation}
where $f$ and $\upsilon$ are the oscillator strength and
thermal kinetic velocity of the ion. $E$ represents the line-center
energy. The line-center optical depth is given by

\begin{equation}
{\rm \tau} = \left(\frac{{n_z}}{{n_Z}}\right)\left(\frac{{n_Z}}{{n_H}}\right)\left(\frac{{n_H}}{{\rm n_e}}\right){n_e\sigma L},
\end{equation}
where ${n_z}/{n_Z}$ is the ionization fraction, ${n_Z}/{n_H}$ is
the relative abundance, and $n_e$ is the electron density, $L$ the path
length through the plasma. For the O VII K$\alpha$ line, $\upsilon$,
$f$, ${n_z}/{n_Z}$, ${n_Z}/{n_H}$, $n_e$ and $L$ are about 100 ${\rm
km\ s}^{-1}$, 0.72, 0.0026, 2$\times 10^{-4}$, 1.1$\ \rm{cm^{-3}}$ and 10
pc respectively. So ${\rm \tau}$ is about 1.9$\times 10^{-3}$
indicating that resonance line scattering is not important for the
O VII K$\alpha$ line. However for the O VIII Ly$\alpha$ line, the
oscillator strength is 0.42 and the ionic fraction is about 0.2
from AtomDB\footnote{http://www.atomdb.org/} which gives ${\rm \tau}$ $\sim$0.3. This value is
somewhat of an underestimate because most of the photons are emitted
in the layer where the O VIII ionization fraction peaks and the
ionization fraction is probably more like 0.5 which increases $\tau$ to
around 0.75.  Thus, the apparent low abundance of O might be
accounted by resonance scattering of the strongest O lines.\\

Two other potential reasons that can cause incorrect O abundance
estimates are: 1) the energy resolution of ACIS detector is too
low to separate different lines in the 0.5 to 1 keV range; 2) one single {\it
vpshock} model may be too simple to fit the X-ray spectrum correctly.
If those are the reasons for the unusual O abundances, the
real abundance uncertainty of Si and Fe should be larger than what
we cited. The next generation X-ray telescopes, such as {\it
Athena}, may help to solve this problem because of their high energy
resolution.\\

\begin{acknowledgements}

HZ and WWT acknowledge support from National Key R\&D Program
of China(2018YFA0404203, 2018YFA0404202) and NSFC (11603039, U1738125). POS acknowledges support
from NASA contract NAS8-03060. The authors thank T. Temim for providing
helpful comments after reading a draft version of this article.

\end{acknowledgements}

\bibliographystyle{apj}

\end{document}